\documentstyle[aps,prb,amstex,multicol,epsfig]{revtex}

\title{Stochastic vortex dynamics in two-dimensional easy-plane ferromagnets:\\
Multiplicative versus additive noise}

\author{Till Kamppeter$^{\ddag}$ and Franz G.\ Mertens$^{\S}$}

\address{Physikalisches Institut, Universit\"at Bayreuth, D-95440 
Bayreuth, Germany}

\author{Esteban Moro$^*$ and Angel S\'anchez$^{\dag}$} 

\address{Grupo Interdisciplinar de Sistemas Complicados (GISC),
Departamento de Matem\'aticas,
Universidad Carlos III de Madrid, E-28911 Legan\'{e}s, Madrid, Spain}

\author{A.\ R.\ Bishop$^{\P}$} 

\address{Theoretical Division and Center for Nonlinear Studies, 
Los Alamos National Laboratory, Los Alamos, New Mexico 87545}

\date{\today}

\newcommand{\pdiff}[2]{\frac{\partial #1}{\partial #2}}

\newcommand{\rK}[1]{\left( #1 \right)}           
\newcommand{\eK}[1]{\left[ #1 \right]}           

\begin{document}

\draft

\maketitle

\begin{abstract}
We study how thermal fluctuations affect the dynamics of vortices 
in the two-dimensional classical, ferromagnetic, 
anisotropic Heisenberg model depending on their
additive or multiplicative character. Using a collective coordinate 
theory, we analytically show that multiplicative 
noise, arising from fluctuations in the local field term of the Landau-Lifshitz
equations, and Langevin-like additive noise both have the same 
effect on vortex dynamics (within a very plausible assumption consistent with 
the collective coordinate approach).
This is a non-trivial result, as multiplicative
and additive noises usually modify the dynamics quite differently.
We also carry out numerical simulations of both versions of the model 
finding that they indeed give rise to very similar vortex dynamics.
\end{abstract}


\begin{multicols}{2}
\narrowtext

\section{Introduction}

In a large class of nonlinear problems arising in physics, chemistry
and biology, coherent, localized excitations often play 
a crucial r\^ole in governing the dynamics of the corresponding systems. 
This is the case, for instance, with solitons, vortices, fronts and many
other solitary wave-like objects found in a wide variety of
low dimensional systems.\cite{book1,book2,book3} Physical situations 
featuring these phenomena are usually described by one of a few ``canonical''
partial differential equations, either integrable, such as the 
one-dimensional sine-Gordon
or the nonlinear Schr\"odinger equations, or non-integrable, such as the
$\phi^4$ one.\cite{Remoissenet} However, those mathematical formulations 
correspond in general to highly idealized physical contexts, in which 
factors like inhomogeneities, fluctuations, external fields, or damping
are not taken into account. Conclusions about stability, 
dynamics, interactions and any other property of coherent excitations 
drawn from those simple descriptions do not necessarily carry over 
when the above ingredients can not be 
neglected. As a consequence, interest on the effect of perturbations 
on solitons and related excitations has grown rapidly since the early 
eighties, motivated by the need of bridging the gap between ideal models 
and real applications. 

One of the most important and universal perturbations of any physical system
is noise in one of its various forms.\cite{vanKampen} Typically, noise 
enters the physics of a system either as temporal
fluctuations of internal degrees 
of freedom, caused by temperature, for instance, or as random variations of
some external control parameter. In order to study the effect of
these fluctuations 
in the system one is interested in, random terms are added to the original
deterministic equations; generally speaking, internal randomness will reflect 
itself in {\em additive} noise terms, while external fluctuations will 
almost always give rise to {\em multiplicative} noise terms. The difference
between the two cases is that additive noise describes fluctuations independent
of the values of the system variables, whereas multiplicative noise relates
to fluctuations whose magnitude is modified by the state of the system. Of 
course, adding one kind of noise or the other to an otherwise deterministic
problem leads in general 
to very different results, and nonlinear coherent excitations 
are not an exception: Thus, for instance, studies of the sine-Gordon
\cite{PV} and $\phi^4$ (Refs.\ \onlinecite{RV,SV}) equations have shown that
large scale (i.e., spatially homogeneous) noise modifies the dynamics of 
solitons in very different ways depending on its additive or multiplicative
character. As another example, we note that 
the difference between additive and multiplicative noise in 
the nonlinear Schr\"odinger equation has also been discussed in Ref.\ 
\onlinecite{Kim1}, where multiplicative noise is associated with 
scattering of solitons by phonons with no creation of new phonons, whereas
additive noise implies creation and anihilation of phonons. However, a 
general discussion of the analogies and differences of both kinds of 
fluctuations is lacking in nonlinear partial differential equations.
Moreover, studies of
noise effects on model systems are often carried out without any reference to 
the physical meaning of the type of term introduced; hence, we believe 
that a physically clear-cut example will help understand the similarities and
differences of additive and multiplicative noise in other cases. 

In this paper, we aim to gain insight into the effects of the two types of
fluctuations by presenting a comparison of the effects of additive and 
multiplicative noise on the dynamics of vortices in two-dimensional (2D) 
easy-plane ferromagnets, as described by the classical, 
anisotropic Heisenberg 
model. The reason we choose this system is that we can justify 
physically in very direct ways the reasons for introducing one or the 
other type of noise in the Landau-Lifshitz equation,
thus making possible the discussion and interpretation
of our results in physical terms. To this end, we will address the problem
both from analytical and numerical viewpoints in order to 
achieve a more complete picture of the two cases. Accordingly, in 
Section 2 we introduce our model, summarize its main features, and 
discuss how noise can be introduced in either form according to 
the physics one has in mind. In Sec.\ 3, we present our analytical 
results, obtained in the framework of a collective coordinate approach.
This approach will allow us to show the surprising result 
that, with a reasonable assumption, very natural in the context of a 
collective coordinate theory,
the two kinds of fluctuations considered lead to the {\em same dynamics}
for a single vortex. Our analytical predictions are tested in Sec.\
4 by means of numerical simulations, which yield a very good agreement 
with the theory thus confirming {\em a posteriori} 
the validity of our assumption. 
Finally, Sec.\ 5 discusses our results and summarizes our main conclusions.
It is important to note that, in principle, a formulation alternative to 
the one presented here is possible in terms of the Hamilton equations,
instead of the Landau-Lifshitz equation. 
We discuss this possibility in an appendix and show that it suffers from 
several problems. 

\section{Model and stochastic perturbations} 

The model we will be working with is 
the 2D anisotropic Heisenberg model with
$XY$- or easy-plane symmetry, given by
\begin{equation} \label{eq:Hamiltonian}
   H=-J \sum_{<m,n>} \eK{S_m^xS_n^x+S_m^yS_n^y+(1-\delta)S_m^zS_m^z},
\end{equation}
where the subindices $x,y$ or $z$ stand for the spin components,
$0<\delta\leq 1$, and $<m,n>$ labels nearest neighbors of a square
lattice. Among its excitations, specially interesting ones are vortices,
that are planar (i.\ e., with null $z$ component) if $\delta\geq 0.297$
and non-planar (i.\ e., with localized $S_z$ structure) if $\delta\leq
0.297$.\cite{Gouvea89,nueva2}.
Such non-planar vortices will be the specific object of our study as reported
in the remainder of the paper; however, the ideas we will be discussing
are general enough to be of interest in other, related contexts where
the system behavior is governed by soliton-like collective excitations.

Physically, this model has many interesting applications:
In the last few years several classes of materials have been found or
fabricated for which magnetic interactions within planes of their crystalline
structure are much stronger than between these planes, and therefore the
magnetic properties are
basically 2D. Materials in these classes
include, for instance, layered magnets (such as Rb$_2$CrCl$_4$), graphite
intercalated compounds (such as CoCl$_2$), magnetic lipid layers (such as
manganese stearate), and high $T_c$ superconductors (see references
in, e.\ g., Ref.\ \onlinecite{Voelkel90}). It is evident that in order 
to model better these materials, one of the factors that has to be 
added to a description in terms of the Heisenberg Hamiltonian is 
fluctuations, which can arise from different origins. Among those, 
thermal noise is of course the most natural source of fluctuations to 
consider in the context of the Heisenberg model: Indeed, 
from the experimental point of view,
insofar as the motion of vortices
has measurable consequences in
inelastic neutron scattering \cite{exp1} and nuclear magnetic resonance
experiments,\cite{exp2} the effects of finite temperature
on vortex dynamics can have signatures in those measurements.

In most cases, thermal effects are studied by adding to the equations 
of motion of the system under consideration an additive noise term 
plus a damping term in order to ensure that the fluctuation-dissipation 
theorem holds. This is simply the familiar Langevin dynamics. 
In our case, our starting (deterministic) point is the Landau-Lifshitz
equation, which reads 
\begin{equation} \label{eq:DFT}
   \frac{d\vec{S}_m}{dt}=
       -\vec{S}_m\times\pdiff{H}{\vec{S}_m},
\end{equation}
where $\vec{S}_m$ is the spin vector at lattice site $m$, and $H$ is the
Hamiltonian, in our case that of the anisotropic Heisenberg
model, Eq.\ (\ref{eq:Hamiltonian}).
The corresponding Langevin dynamics equation for our model is obtained 
by adding damping and additive noise, which yields
\begin{equation} \label{eq:Gilbert3}
   \frac{d\vec{S}_m}{dt}=
       -\vec{S}_m\times\frac{\partial H}{\partial\vec{S}_m}
          -\epsilon\vec{S}_m\times\frac{d\vec{S}_m}{dt}
          +\vec{\eta}_m(t).
\end{equation}
The second term on the rhs of Eq.\ (\ref{eq:Gilbert3}) is the damping
term. Following Refs.\ \onlinecite{Thiele73}, \onlinecite{Thiele74}
and \onlinecite{Huber82} we have chosen for simplicity Gilbert
damping,\cite{nota1} chiefly because it is isotropic in contrast to
the Landau-Lifshitz damping.\cite{Iida62} The last term in the rhs of
Eq.\ (\ref{eq:Gilbert3}) is the noise term, given by a Gaussian white
noise with statistics defined by
\begin{mathletters}
\label{ggfg}
\begin{eqnarray} \label{eq:Noise_mean}
   \langle\eta_{m\alpha}(t)\rangle &=& 0, \\
\label{eq:Noise_variance}
   \langle\eta_{m\alpha}(t)\eta_{n\beta}(t')\rangle &=&
      D \delta_{mn}
      \delta_{\alpha\beta}\delta(t-t'),
\end{eqnarray}
\end{mathletters}
where $D=2\epsilon k_{\rm B} T$ is the diffusion constant and
$\alpha,\,\beta = 1,\, 2,\, 3$ denote cartesian coordinates.  It is
important to note that, strictly speaking, the three Eqs.\ 
(\ref{eq:Gilbert3}) do not represent Langevin equations, because {\em
  all} the components of $d\vec{S}_m/dt$ appear in each equation due
to the cross products.  To properly introduce the noise, one first has
to group all the time derivatives on the lhs of the equation, and only
then one can add independent white noise terms, say $\vec{\rho}_m$,
for each spin component. In Refs.\ \onlinecite{Till} and
\onlinecite{Till3} it was shown that such a procedure leads to a
different Langevin equation, in which the noise term $\vec{\eta}_m$
must be replaced by $\vec{\eta}_m=\vec{\rho}_m-
\epsilon(\vec{S}\times\vec{\rho}_m)$. However, as discussed in Refs.\ 
\onlinecite{Till} and \onlinecite{Till3}, the correction is of the
order of $\epsilon^2$, and taking into account that in the simulations
values of $\epsilon$ of the order of $10^{-3}$ are used, the
correction factor can be neglected. I.~e.\ we take Eq.\ 
(\ref{eq:Gilbert3}) as Langevin equation, containing {\em purely
  additive noise}.

What we have discussed is the usual way to introduce thermal
fluctuations in any model. Nevertheless, in our case this Langevin
approach suffers from the drawback that it enlarges the length of the
spins, which in the original Heisenberg model is fixed to
$|\vec{S}|=1$. In the simulations of Ref.~\onlinecite{Till3}, this
unphysical effect was suppressed by renormalizing the length of every
spin back to its original length at every time step. A more detailed
discussion of the implementation of this constraint can be found in
Ref.~\onlinecite{Till}. At this point, we were not fully satisfied
with this solution, and looked for another way to study thermal
effects which preserve the spin length {\em exactly}.  With this
motivation, we now propose to introduce a noise term {\em
  multiplicatively}, according to
\begin{equation} \label{eq:mult1}
\frac{d\vec{S}_m}{dt}=
-\vec{S}_m\times\left[\frac{\partial H}{\partial \vec{S}_m}+\vec{h}_m(t)\right]
-\epsilon\vec{S}_m\times\frac{d\vec{S}_m}{dt}.
\end{equation}
The term $\vec{h}_m(t)$ is again a set of independent Gaussian white noises,
but now they represent fluctuations in the local field, $\partial H/
\partial \vec{S}_m$, in which the spin $\vec{S}_m$ precesses. In fact, this 
is a natural way to introduce the effect of thermal fluctuations, as the 
local field is the only way through which the spin $\vec{S}_m$ can feel 
any changes in its environment, due to those thermal fluctuations or for
any other reason. Thus, the random term accounts for the interaction of 
the spin degrees of freedom with phonons, magnons, 
and any other excitation thermally generated. In addition, this fashion 
of introducing the noise has the property that Eq.\ (\ref{eq:mult1})
exactly preserves the spin length, hence there is no need for
corrections as in the additive case. We note that a similar term has been 
considered by Garanin, \cite{garanin} who proposed it in order to derive
an all-temperature theory from the corresponding Fokker-Planck equation,
obtaining a so-called Landau-Lifshitz-Bloch equation. Our purpose now is
to use multiplicative noise
in order to understand the influence of finite temperature on 
vortex dynamics, comparing the results 
with those arising from the usual Langevin
approach discussed above. 

There is an important question that deserves discussion before
proceeding to the study of vortex dynamics, namely the correct way to
interpret the stochastic partial differential equations
(\ref{eq:Gilbert3}) and (\ref{eq:mult1}). The first one contains only
additive noise, which implies that Ito or Stratonovi\v c
interpretations coincide,\cite{vanKampen,gard} and therefore there is
no problem in that case. As for the second one, being multiplicative,
we do have to specify our interpretation of the equation.  In
principle, when thinking of thermal excitations interacting with the
spins, we would have to associate with them a finite correlation time
which would lead to a colored noise term. Taking white noise means
taking the limit of zero correlation time, and therefore it is
necessary to interpret Eq.\ (\ref{eq:mult1}) in the Stratonovi\v c
sense. Another reason for us to stay with this interpretation is that
the spin modulus is conserved; it can be seen that Ito calculus leads
to an exponential decrease of the modulus, with a damping time
proportional to the damping $\epsilon$ in Eq.\ (\ref{eq:mult1}).  In
addition, in the Stratonovi\v c interpretation, Garanin,\cite{garanin}
and even earlier Brown,\cite{Brown} showed\cite{nota2} that the
stationary solution to the Fokker-Planck equation corresponding to
Eq.\ (\ref{eq:mult1}) is the Boltzmann factor with Hamiltonian
(\ref{eq:Hamiltonian}), indicating that Eq.\ (\ref{eq:mult1}) indeed
represents the dynamics of our model at finite temperatures.
Interestingly, it is not difficult to show that the Boltzmann factor
is {\em not} a stationary solution to the Fokker-Planck equation for
the additive noise Langevin problem, Eq.\ (\ref{eq:Gilbert3}), unless
further assumptions are made, including the constraint of constant
spin modulus\cite{TillPhD}.

\section{Analytical results}

As stated in the Introduction,
our approach to the problem of vortex dynamics
will be both analytical
and numerical. In this section, we 
first derive equations of motion for the vortex
center $\vec{X}(t)$ both for the additive and for the multiplicative noise
cases, and
afterwards we compare with numerical simulations for
our model, i.\ e., with results from numerical integration of
Eqs.\ (\ref{eq:Gilbert3}) and (\ref{eq:mult1}).

Our analytical approach to the stochastic dynamics of vortices
begins by taking the continuum limit of Eqs.\ (\ref{eq:Gilbert3}) 
and (\ref{eq:mult1}), which are much more difficult to deal with in 
a lattice formulation. This is a good approximation provided that the 
localized $S_z$-structure
spans many lattice sites, as typically occurs in practice (except if the 
anisotropy parameter is chosen close to the critical value 
$\delta=0.297$). The next
step is to use a collective coordinate theory to analyze the vortex 
dynamics. 
(see Ref.\ \onlinecite{review} for a recent review on collective 
coordinate approaches). Within this procedure, one assumes that 
the shape of the excitation under consideration, in our case a vortex, 
is not modified by the perturbation for a large range of perturbation 
types (this is a 
very general, widely applicable\cite{review} approach),
i.e., in our case noise and damping terms, and
that only the dynamics of its center is modified by these extra
terms. 
The vortex motion is then introduced by the {\em travelling wave 
Ansatz} $\vec{S}(\vec{r},t)=\vec{S}(\vec{r}-
\vec{X}(t))$, where $\vec{S}(\vec{r})$ describes the static 
vortex shape. Unfortunately, such a simple 
approach 
(first proposed for magnetic domains by Thiele\cite{Thiele73,Thiele74}) 
is not enough to describe the vortex dynamics, as was found in
Ref.\ \onlinecite{Mertens96}. In that work, Mertens {\em et al.} 
developed a generalization of the collective coordinate theory 
in which the vortex shape is allowed to 
depend on the velocity $\dot{\vec{X}}$ and,
in general, also on higher order derivatives
of $\vec{X}(t)$. The corresponding {\em generalized travelling 
wave An\-satz} is
\begin{equation} \label{eq:trav_wave}
\vec{S}(\vec{r},t)=\vec{S}(\vec{r}-\vec{X},\dot{\vec{X}},\ddot{\vec{X}},
\ldots,\vec{X}^{(n)})  ,
\end{equation}
which yields an $(n+1)$-th order differential equation for $\vec{X}(t)$. 
However, as
discussed in Ref.\ \onlinecite{Mertens96}, 
in the case of non-planar vortices only the odd-order equations represent
self-consistent valid approximations; and it turns out
that the third-order equ\-ation is sufficient to describe
accurately all simulations without damping
\cite{Mertens96}. Therefore, in this paper we use the {\em Ansatz}
(\ref{eq:trav_wave}) with $n=2$ and apply it to the general case,
which includes damping and noise. 

The continuum versions of Eqs.\ (\ref{eq:Gilbert3}) and (\ref{eq:mult1})
read, respectively
\begin{equation} \label{eq:Gilbert4}
   \frac{d\vec{S}}{dt}=
       -\vec{S}\times\frac{\delta H}{\delta\vec{S}}
          -\epsilon\vec{S}\times\frac{d\vec{S}}{dt}
          +\vec{\eta}(\vec{r},t).
\end{equation}
and
\begin{equation} \label{eq:mult2}
\frac{d\vec{S}}{dt}=
-\vec{S}\times\left[\frac{\delta H}{\delta \vec{S}}+\vec{h}(\vec{r},t)\right]
-\epsilon\vec{S}\times\frac{d\vec{S}}{dt}.
\end{equation}
To obtain the equations for the collective coordinate, instead of using 
the Hamiltonian procedure described in Ref.\ \onlinecite{Mertens96},
we follow a much more direct approach, which we have already used for
the additive noise (see a preliminary report in Refs.\ 
\onlinecite{Till} and \onlinecite{Till2}):
We begin with Eq.\ (\ref{eq:Gilbert4}) 
and multiply it by $\vec{S}\cdot(\partial\vec{S}/ 
\partial X_i)\times$, where $X_i$ is the $i$-th component of the vortex 
center position. The contributions of all terms in the rhs of Eq.\ 
(\ref{eq:Gilbert4}) are
\begin{mathletters}
\label{add}
\begin{eqnarray}
\label{add1}                                                   
\nonumber
-\vec{S}\cdot\rK{\pdiff{\vec{S}}{X_i}\times
      \eK{\vec{S}\times\frac{\delta H}{\delta \vec{S}}}} &=&
  -S^{2}\frac{\delta H}{\delta \vec{S}}\cdot\pdiff{\vec{S}}{X_i}= \\
& & =-S^{2}\pdiff{{\cal H}}{X_i},\\
\label{add2}
    \epsilon\vec{S}\cdot\eK{\pdiff{\vec{S}}{X_i}\times
         \rK{\vec{S}\times\frac{d\vec{S}}{dt}}} &=&
      \epsilon S^{2}\pdiff{\vec{S}}{X_i}\cdot\frac{d\vec{S}}{dt},\\
\label{add3}
\vec{S}\cdot\rK{\pdiff{\vec{S}}{X_i}\times\vec{\eta}}
      &=&\rK{\vec{S} \times \pdiff{\vec{S}}{X_i}} \cdot \vec{\eta}.
\end{eqnarray}
\end{mathletters}
where ${\cal H}$ is the Hamiltonian density in the continuum limit 
of Eq.\ (\ref{eq:Hamiltonian}).
According to our ansatz we
insert in Eq.\ (\ref{add2})
\begin{equation} \label{eq:dsdt}
   \frac{d\vec{S}}{dt}=
   \pdiff{\vec{S}}{X_j}\dot{X}_j+
      \pdiff{\vec{S}}{\dot{X}_j}\ddot{X}_j+
      \pdiff{\vec{S}}{\ddot{X}_j}\raisebox{0.3mm}{$\dddot{X}_j$}  .
\end{equation}
The lhs of Eq.\ (\ref{eq:Gilbert4}) is dealt with in the same way as 
just described for the rhs. 
By collecting the results,
integrating over $\vec{r}$, and dividing by $S^2$, we obtain the
same third-order equation as that in Refs.\ \onlinecite{Till,Till3}, 
and \onlinecite{Till2}:
\begin{multline} \label{eq:3rd-ord_damp}
   \rK{{\bf A} + {\bf a}}\dddot{\vec X} +
      \rK{{\bf M} + {\bf m}}\ddot{\vec{X}} +
      \rK{{\bf G} + {\bf g}}\dot{\vec{X}} =\\ =
   \hat{\bf A}\dddot{\vec X} +
      \hat{\bf M}\ddot{\vec{X}} +
      \hat{\bf G}\dot{\vec{X}} =
   \vec{F}  +
   \vec{F}^{\rm add} .
\end{multline}
The terms in Eq.\ (\ref{eq:3rd-ord_damp}) are as follows: The tensors denoted
by capital letters come from the lhs of Eq.\ (\ref{eq:Gilbert4}),
and their expressions are,
for the {\em gyrotensor} ${\bf G}$,
\begin{equation} \label{eq:Gij}
   G_{ij} =
      S^{-2}\int\!d^2r\,\vec{S}\cdot\rK{\pdiff{\vec{S}}{X_i}\times
                                        \pdiff{\vec{S}}{X_j}}
\end{equation}
for the {\em mass tensor} ${\bf M}$, 
\begin{equation} \label{eq:Mij}
   M_{ij} =
      S^{-2}\int\!d^2r\,\vec{S}\cdot\rK{\pdiff{\vec{S}}{X_i}\times
                                        \pdiff{\vec{S}}{\dot X_j}}
\end{equation}
and for the {\em third-order gyrotensor} ${\bf A}$,
\begin{equation} \label{eq:Aij}
   A_{ij} =
      S^{-2}\int\!d^2r\,\vec{S}\cdot\rK{\pdiff{\vec{S}}{X_i}\times
                                        \pdiff{\vec{S}}{\ddot X_j}}
\end{equation}
The tensors denoted by small letters come from the Gilbert damping term;
as can be seen from Eq.\ (\ref{eq:3rd-ord_damp}), they contribute to all 
orders, and they are given by 
\begin{equation}
   g_{ij} =
      \epsilon\int\!d^2r\,\pdiff{\vec{S}}{X_i}\cdot
                  \pdiff{\vec{S}}{X_j} 
 , \label{eq:D1ij}
\end{equation}
\begin{equation}
   m_{ij} =
     \epsilon\int\!d^2r\,\pdiff{\vec{S}}{X_i}\cdot
                \pdiff{\vec{S}}{\dot X_j} 
 ,  \label{eq:D2ij}
\end{equation}
\begin{equation}
   a_{ij}
    =   \epsilon\int\!d^2r\,\pdiff{\vec{S}}{X_i}\cdot
                  \pdiff{\vec{S}}{\ddot X_j} 
 . \label{eq:D3ij}
\end{equation}
Finally, the force terms are 
\begin{equation} \label{eq:F}
   F_i = -\int\!d^2r\, \pdiff{{\cal H}}{X_i}  ,
\end{equation}
and
\begin{equation} \label{eq:coll_noise_1}
   F_i^{\rm add} =
   \frac{1}{S^2} \int d^2r \rK{\vec{S} \times \pdiff{\vec{S}}{X_i}}
      \cdot\vec{\eta}(\vec{r},t).
\end{equation}
The key to achieve a complete understanding of the vortex dynamics 
as described
by Eq.\ (\ref{eq:3rd-ord_damp}),
is to know the mean $\langle F_i^{\rm add} \rangle$ and
the variance ${\rm Var}(F_i^{\rm add})$. 
The mean is easily shown to be zero, whereas
for the correlation functions,
by using the 
continuum version of Eq.\ (\ref{eq:Noise_variance}) we obtain
\begin{multline} \label{eq:Fst_var}
   \langle F_i^{\rm add}(t) F_j^{\rm add}(t') \rangle
    = \\D \delta_{ij} \delta(t-t') \int d^2 r
\frac{\partial\vec{S}}{\partial X_i}\cdot\frac{\partial\vec{S}}{\partial X_j}.
\end{multline}

We now turn to the multiplicative noise case, Eq.\ (\ref{eq:mult2}).
The calculations are mostly
the same as described above, except that the contribution in Eq.\ 
(\ref{add3}) is now substituted by the term coming from the multiplicative 
noise, which reads
\begin{equation}
\vec{S}\cdot\rK{\pdiff{\vec{S}}{X_i}\times(\vec{S}\times\vec{h})}=
S^2\frac{\partial\vec{S}}{\partial X_i}\cdot\vec{h}.
\label{mult2}
\end{equation}
As a consequence, only the stochastic term in  Eq.\ (\ref{eq:3rd-ord_damp})
is modified: The new stochastic force is 
\begin{equation} \label{eq:multnoise}
   F_i^{\rm mult} =
   \frac{1}{S^2} \int d^2r \pdiff{\vec{S}}{X_i}\>
      \cdot\vec{h}(\vec{r},t),
\end{equation}
which is in principle different from what we obtained for the additive 
case, $F^{\rm add}$. However, when we evaluate the first moments, 
we find again that the mean of $F^{\rm mult}$ is zero. Furthermore, 
to compute the variance,  we need to evaluate 
\begin{multline} \label{eq:Fmul_var}
   \langle F_i^{\rm mult}(t) F_j^{\rm mult}(t') \rangle =\\
\langle\int d^2 r\int d^2 r'\pdiff{{S}_{\alpha}}{X_i}
\pdiff{{S}_{\beta}}{X_j}h_{\alpha}(\vec{r},t)h_{\beta}(\vec{r}',t')\rangle.
\end{multline}
where summation over $\alpha$ and $\beta$ is implied.
At this point, it is important to notice that the noise is multiplicative,
and therefore in principle we cannot take the spin fields out of the 
average over realizations of the noise. We stress that this is not a
problem in the case of additive noise, and that we indeed proceeded 
that way to obtain the expression for the variance of $F^{\rm add}$
given in Eq.\ (\ref{eq:Fst_var}). This is so because the additive 
character of the noise implies that the spin fields and the noise 
are uncorrelated. However, one can not simply apply the same argument
to the calculation in Eq.\ (\ref{eq:Fmul_var}), because when the noise
is multiplicative it is not clear whether the spin fields and the noise
are correlated or not. In this situation, in principle 
we can not exactly evaluate
the variance of $F^{\rm mult}$, but we can make the following 
approximation: At least {\em for small noise}, we can substitute the
spin fields in Eq.\ (\ref{eq:Fmul_var}) by the deterministic expression
for the vortices, assuming reasonably that the corrections induced by 
the noise will be of the order of the noise strength and that their 
contribution to the variance would be two orders higher in the noise 
strength than that of the deterministic part, and hence, negligible. 
If we do so, we can then take the spin fields out of the average, 
finding
\begin{multline} \label{eq:Fmultvar2}
   \langle F_i^{\rm mult}(t) F_j^{\rm mult}(t') \rangle
    = \\ D \delta(t-t') \int d^2 r
\frac{\partial\vec{S}}{\partial X_i}\cdot\frac{\partial\vec{S}}{\partial X_j},
\end{multline}
i.e., {\em the variance of the stochastic force due to the multiplicative
noise is the same as that of the additive noise}. 

We stress that the assumption leading to the above, unexpected result, is
very natural within a collective coordinate approach such as the one we 
are using here. To understand this, recall that the main hypothesis of a 
collective coordinate theory is that {\em the shape of the excitation 
under study remains mostly unaffected by the perturbations, and only its
position and possibly a few other collective variables change due to them}.
\cite{review} This assumption amounts to a drastic reduction in the 
degrees of freedom of the system: From an infinite number of them in 
the continuum equations to two for the center dynamics, in the present
case. Physically, it is equivalent to neglect the contribution of the 
magnons\cite{nota4}
excited by the perturbations to the vortex shape; and this 
approximation is made already at the very beginning, in order to obtain 
Eq.\ (\ref{eq:3rd-ord_damp}). In view of this, it would not be reasonable
to keep the exact fields $\vec{S}$, containing the contribution of the
phonon degrees of freedom, in Eq.\ (\ref{eq:Fmul_var}), and therefore 
we carry out the calculation with the deterministic vortex shape, 
obtaining the result (\ref{eq:Fmultvar2}).\cite{nota3} Thus, {\em consistently 
within our collective coordinate approach}, we have shown that the 
mean and the
variance of the stochastic forces induced by the multiplicative and 
the additive noises are the same (higher-order moments 
may still differ). As a consequence, the dynamics of 
a vortex is predicted to be the same under the influence of each 
type of noise, which is a surprising result in view of the 
general result that their effects are very different. 

Once we have shown that both stochastic forces are equal, then 
Eq.\ (\ref{eq:3rd-ord_damp}) has to be the description of the vortex
dynamics under the two types of noise. This equation was already 
studied and solved in Refs.\ \onlinecite{Till3} and
\onlinecite{Till2}, and we will 
only summarize here what we need for our present purposes. 
The classical spin is constrained to have a fixed magnitude which we
set to unity. Therefore, the integrals
on the rhs of Eqs.\ (\ref{eq:Gij}) through (\ref{eq:D3ij})
were calculated using canonical
fields $\phi=\arctan(S_y/S_x)$ and $\psi=S_z$ for the spin vector:
\begin{equation} \label{eq:S_m_phi}
\hspace*{-1.2cm}   \vec{S} = \sqrt{1-\psi^2}\cos\phi\,\vec{e}_x
           + \sqrt{1-\psi^2}\sin\phi\,\vec{e}_y
           + \psi\,\vec{e}_z.
\end{equation}
Now, the
explicit calculation of all the integrals is only possible if
the {\em dynamic} structure of the vortex is known. In this respect,
in Ref.\ \onlinecite{Mertens96} it was shown that the core region of
the vortex contributes very little [except to Eq.\ (\ref{eq:Gij})];
the dominant contributions stem from the outer region, if the system
size is large enough. A vortex in the center of a circular system with
radius $L$ and free boundary conditions has the following structure 
in the outer region
which was confirmed by simulations:\cite{Mertens96}
\begin{equation} \label{eq:dyn_phi_psi}
   \phi=\phi_0+\phi_1+\phi_2  , \quad
   \psi=\psi_0+\psi_1+\psi_2
\end{equation}
with
\begin{equation} \label{eq:static_structure}
            \phi_0 = q\tan^{-1}\frac{x_2}{x_1},
\end{equation}
\begin{equation}
      \phi_1 = p(x_1\dot{X}_1+x_2\dot{X}_2),
\end{equation}
\begin{equation}
            \phi_2 = \frac{q}{8\delta}\log \frac{r}{eL}(x_2\ddot{X}_1-x_1\ddot{X}_2),
\end{equation}
\begin{equation}
            \psi_0 \sim p\sqrt{\frac{r_{v}}{r}}\exp(-r/r_{v}),
\label{jolin}
\end{equation}
\begin{equation}
             \psi_1 = \frac{q}{4\delta r^2}(x_2\dot{X}_1-x_1\dot{X}_2),
\end{equation}
and
\begin{equation}
\label{eq:static_structure2}
            \psi_2 = \frac{p}{4\delta}(x_1\ddot{X}_1+x_2\ddot{X}_2).
\end{equation}
Here $q=\pm 1$ is the vorticity and $p=\pm 1$ is the polarization,
which determines to which side the out-of-plane structure of the
vortex points. In addition, $r_v=[(1-\delta)/\delta]^{1/2}/2$ characterizes
the radius of the vortex core. 
Straightforward integrations then yield expressions for the tensors
$\hat{\bf G}$, $\hat{\bf M}$, and $\hat{\bf A}$,
as well as for the forces, which can afterwards be inserted in 
Eq.\ (\ref{eq:3rd-ord_damp}) for the vortex motion. Finally, this 
equation is linear except for the force $\vec{F}(\vec{X})$, which can 
in turn be linearized by expanding around the mean trajectory. 
Subsequently, the equation can be solved by means of a Green's function
approach and the so obtained solution can be used to calculate analytically
the variances of the vortex trajectory {\bf X}$(t)$. These variances 
are proportional to the effective vortex diffusion constant
\begin{equation}
\label{dv}
D_v\simeq D\pi \left\{\log \frac{L}{a_c} + C(a_c)\right\},
\end{equation}                               
where $a_c$ is chosen of the order of $r_v$, and the unknown 
constant $C$ stems from the core region.               
As in this work our main interest is to show
that the dynamics under additive and multiplicative noises is the 
same, we will not dwell any further in the solution of 
Eq.\ (\ref{eq:3rd-ord_damp}); the interested reader is referred to Refs.\ 
\onlinecite{Till3} and \onlinecite{Till2} for details.

\section{Numerical results} 

Our numerical simulations
begin with one vortex with its center located at a distance $R_0$ from
the middle of a circularly shaped square lattice with a
radius of $L$ lattice constants. We use free boundary conditions to 
produce an
image antivortex which leads to a radial force on our
vortex.\cite{Mertens96,Schnitzer96a}
The initial spin configuration is determined from an iterative program
which produces a discrete vortex structure on the lattice.
\cite{Schnitzer96a} In this way we avoid the radiation of spin waves
which would appear during the early time units if we use a continuum
or other
approximation for the vortex shape. The parameters we used in our 
simulations were: $\delta=0.1$, for the radius of the out-of-plane 
structure of the vortex to be large enough to avoid severe discreteness
effects; system radius $L=24$ for the vortex to have enough space 
to evolve far from the boundaries; and initial distance from the 
circle center $R_0=10$. 

For the time integration of the Landau-Lifshitz equation we use the
discrete version of (\ref{eq:Gilbert3}). Consistently with our 
interpretation of all the equations in the sense of Stratonovi\v c, 
we use the Heun method to integrate in time.\cite{36,37}
We explicitly take into account the constraint $\vec{S}^2=1$
by means of a Lagrange parameter, as discussed in  
Ref.\ \onlinecite{niels} (see also Ref.\ \onlinecite{Till3}).
To find a proper damping constant
we checked the time dependence of the system energy using different damping
constants for $L=24$ and $T=0.02$ (in dimensionless units).
The energy at $t=0$ is
the same as for $T=0$ and $\epsilon=0$ because the noise is introduced
with the first time step of the simulation.
The energy then rises and saturates to a value
independent of $\epsilon$, but for $\epsilon > 8 \times 10^{-3}$ the
energy decreases slowly after saturation. The saturation time gets
longer with lower $\epsilon$, for $\epsilon \geq 2 \times 10^{-3}$ we
achieve acceptable saturation times $< 300$ (in units of $\hbar/(JS)$).
We have always made a pre-run of length $t_0>300$ prior to beginning
the evaluation of the simulation data.
Finally, our simulations consisted of numerical integrations up
to times $t=4\,000$ (note in
this regard that this takes 10 days CPU time on a 433-MHz-Digital-Alpha workstation
for averages over 100 runs) because this is larger than the characteristic
time given by $5/\epsilon$ for the damping in the trajectories.
Finally, 
the difference between the energy without temperature and the saturation
energy with temperature must be the thermal energy. We computed the
mean thermal energy per spin at several temperatures and it agreed with
$f/2 \times k_{\rm B} T$ up to $T=0.9$ (for comparison, we note that 
the Kosterlitz-Thouless transistion temperature is about 0.8 in our units),
$f$ being the number of degrees
of freedom per spin. This is far above the temperatures we will discuss below. 
We have to mention that for $T\gtrsim 0.1$ there is a more complex 
phenomenology because the thermal noise does
not only induce a diffusive motion of 
the vortices but it can also flip their out-of-plane structure, 
correspondingly changing their direction of motion, and also 
nucleate additional vortex-antivortex pairs. In this work we 
stay away from this regime which has been considered in 
Ref.\ \onlinecite{flips}.

\begin{figure}[H]
\noindent
\epsfig{file=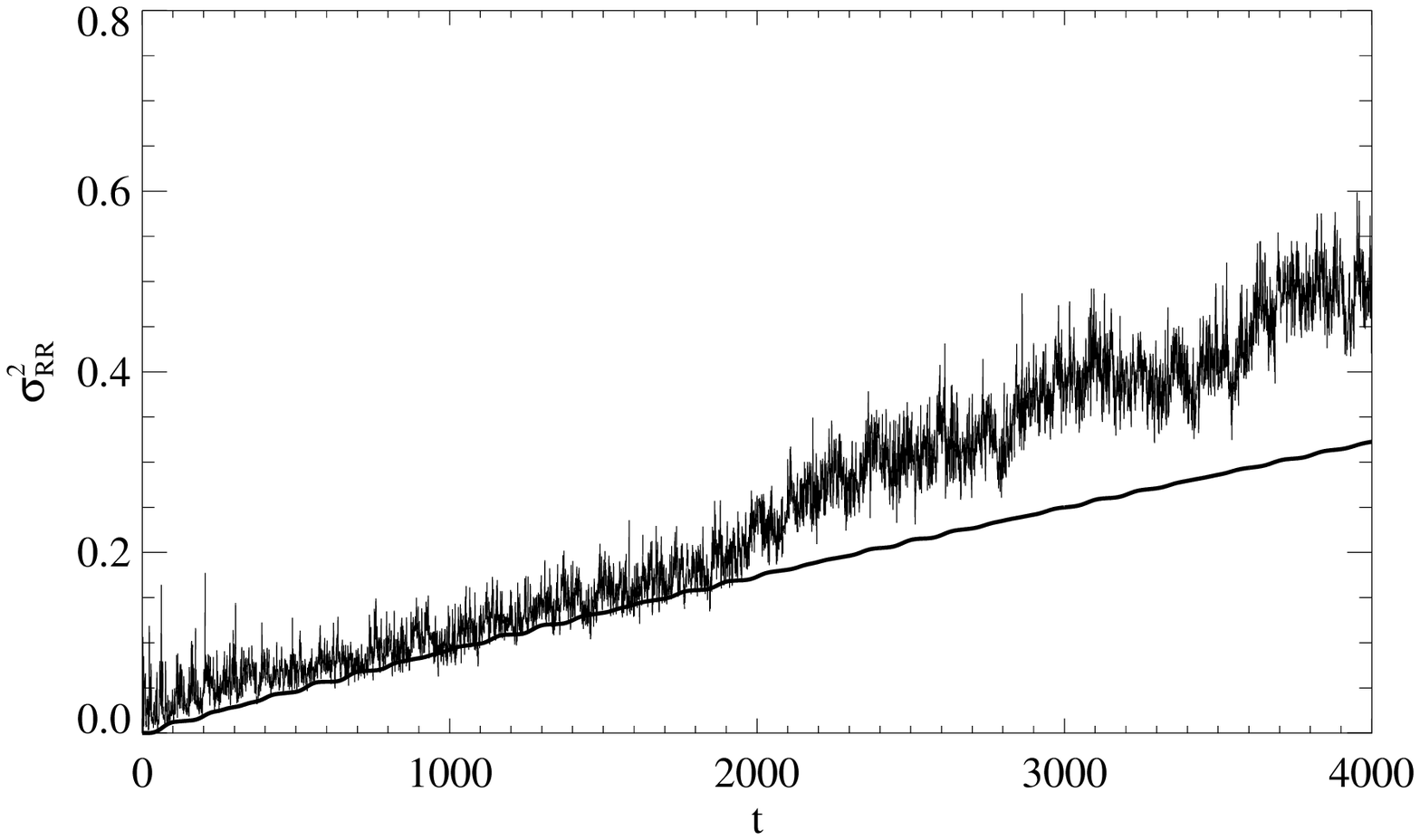, width=3.0in}
\epsfig{file=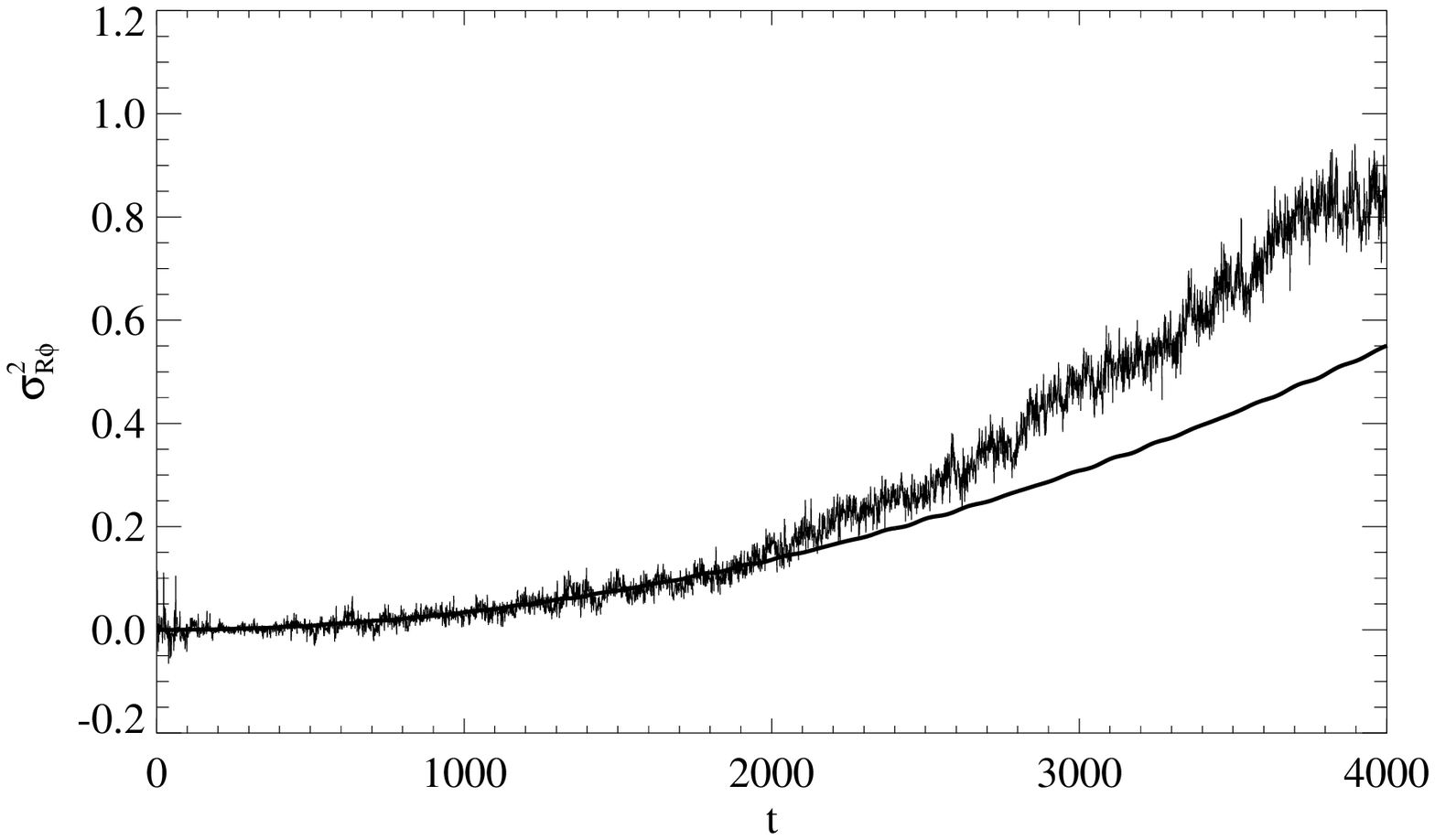, width=3.0in}
\epsfig{file=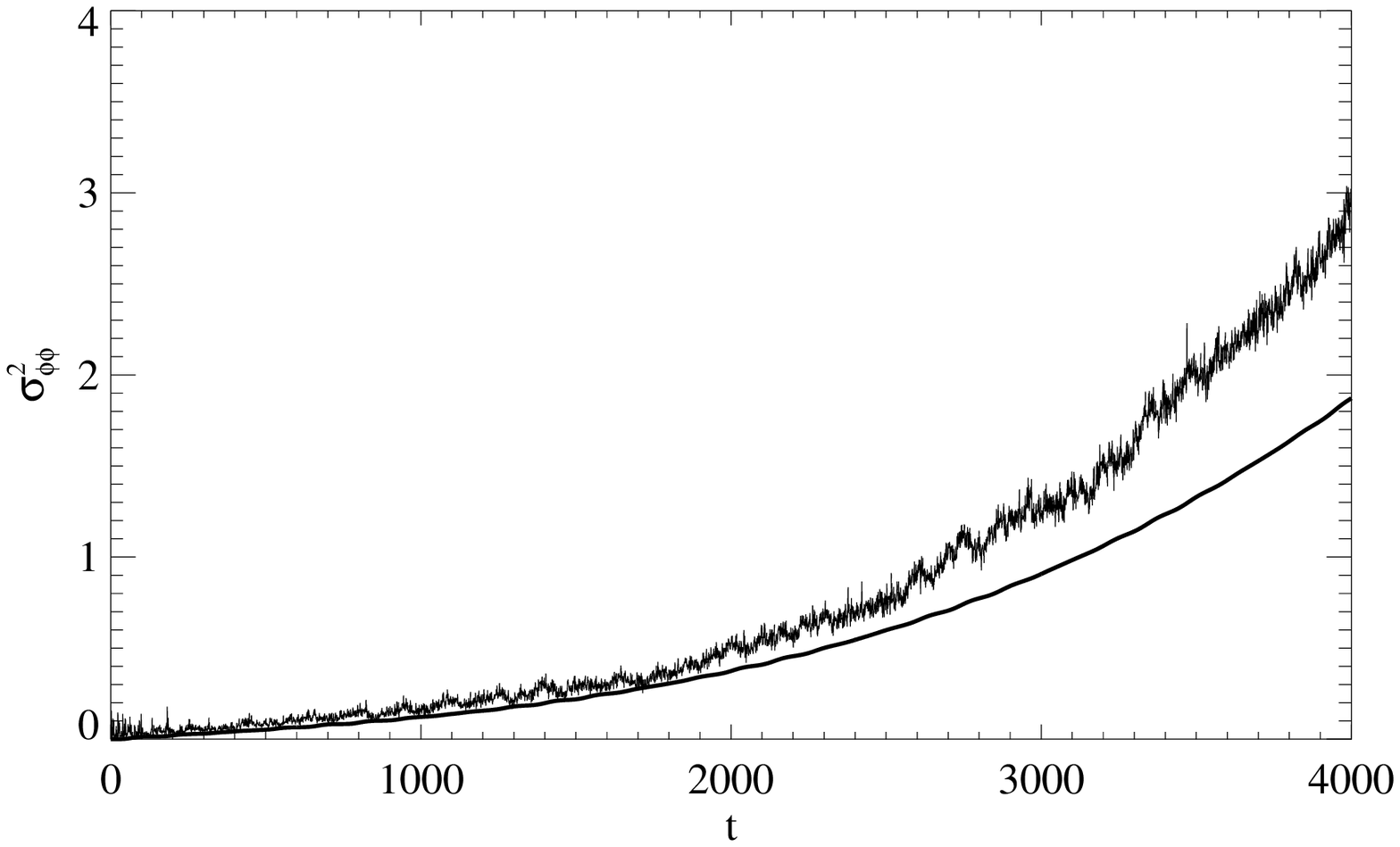, width=3.0in}
\caption{Variances of the vortex trajectory for Langevin dynamics with
{\em additive noise}. The temperature is $T=0.03$; other parameters are 
given in the text.
{}From top to bottom, shown are the variance of the radial coordinate,
$\sigma_{RR}^2=\langle R^2\rangle -\langle R\rangle^2$, $\sigma_{R\phi}^2$,
the off-diagonal elements of the variance matrix, and $\sigma_{\phi\phi}^2$,
the variance of the azimuthal coordinate. In all
three cases the lower line is the theoretical prediction with
the vortex diffusion constant chosen in order to make the analytical
curve lie just below the simulated data.}
\label{fig1}
\end{figure}

\begin{figure}[H]
\noindent
\epsfig{file=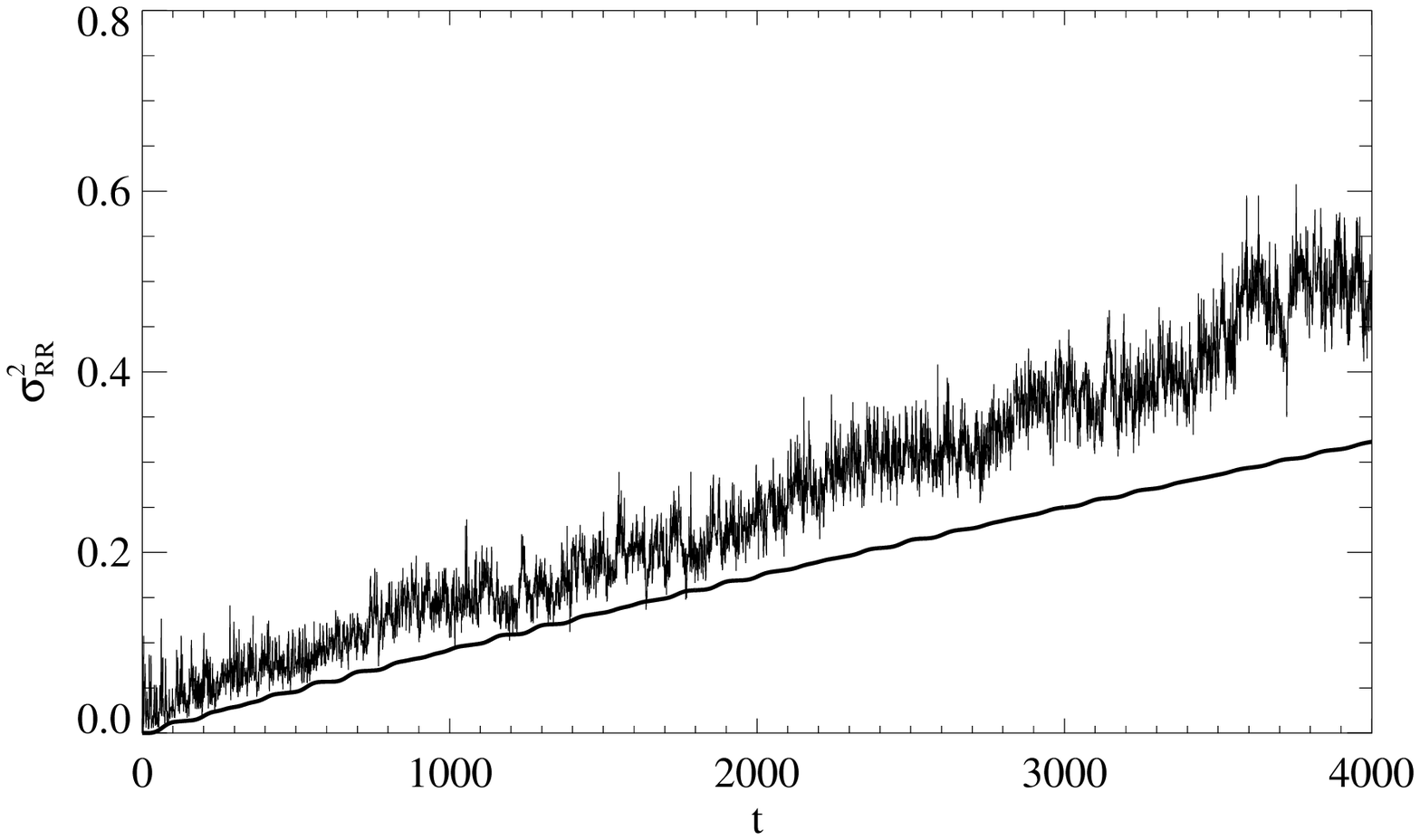, width=3.0in}
\epsfig{file=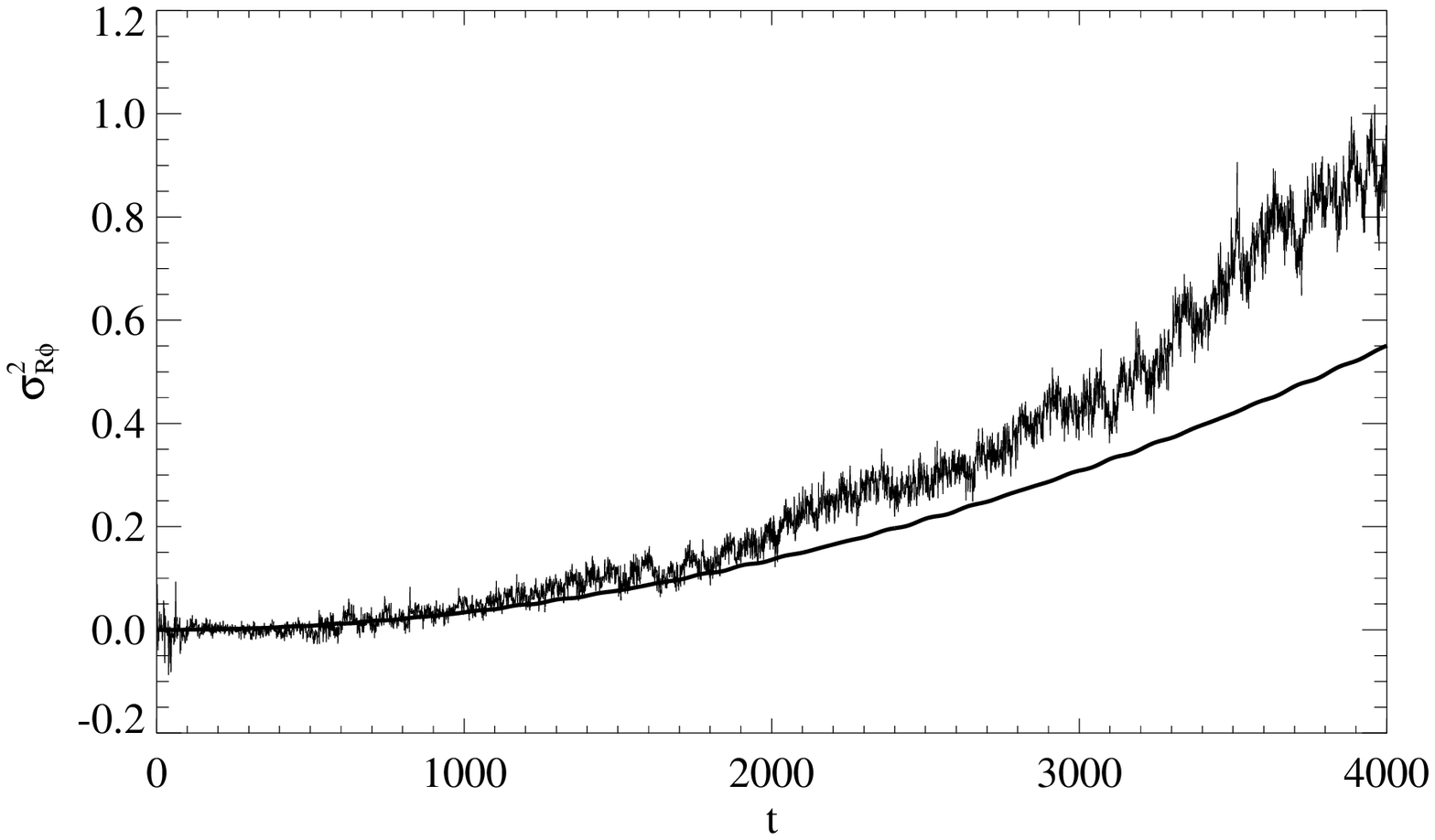, width=3.0in}
\epsfig{file=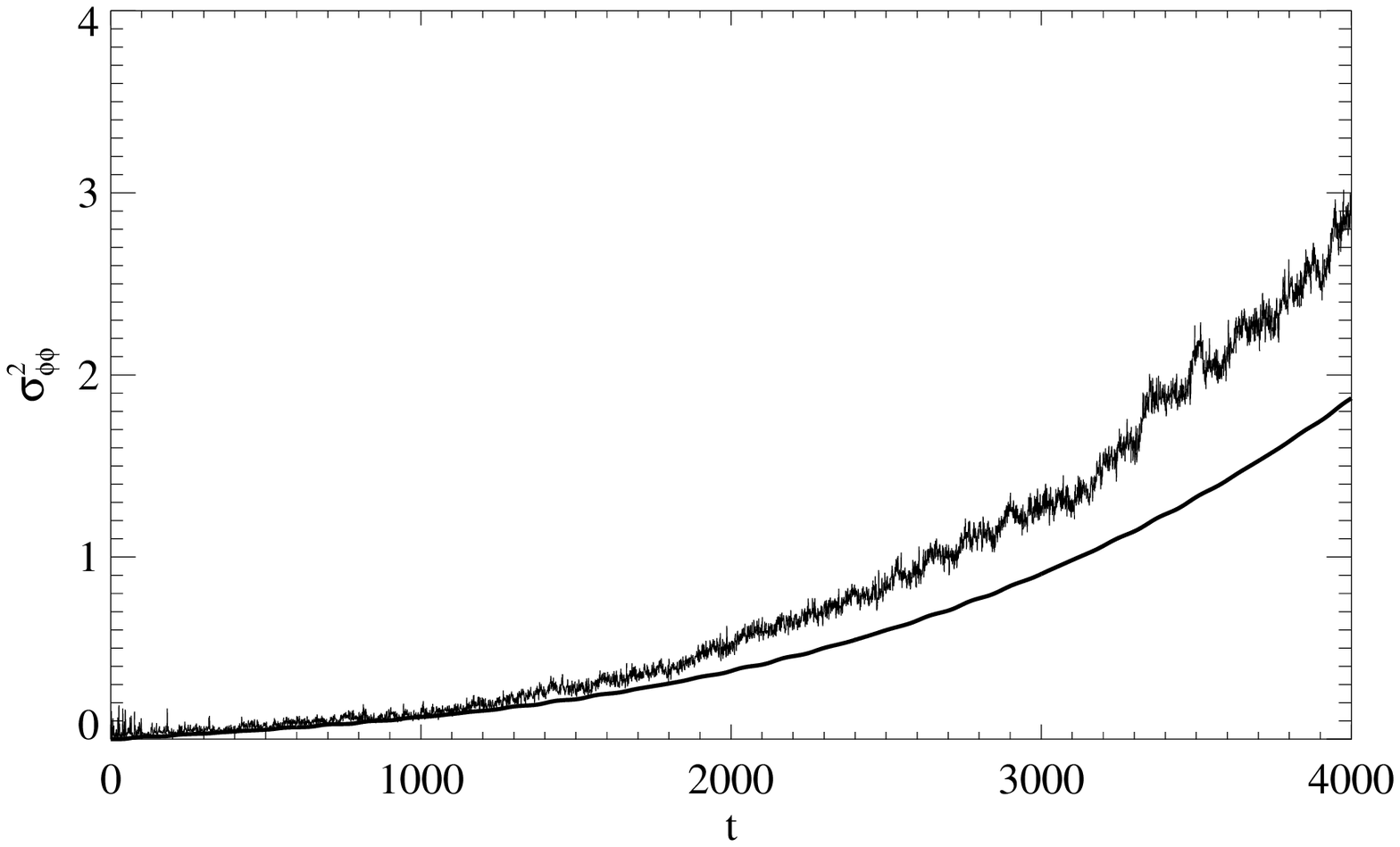, width=3.0in}
\caption{Variances of the vortex trajectory for the case of
{\em multiplicative} noise; parameters are the
same as in Fig.\ \ref{fig1}.
{}From top to bottom, shown are the variance of the radial coordinate,
$\sigma_{RR}^2=\langle R^2\rangle -\langle R\rangle^2$, $\sigma_{R\phi}^2$,
the off-diagonal elements of the variance matrix, and $\sigma_{\phi\phi}^2$,
the variance of the azimuthal coordinate. In all
three cases the lower line is the theoretical prediction with
the vortex diffusion constant fitted to the data.}
\label{fig2}
\end{figure}

The outcome of our numerical simulations can be summarized by saying 
that they fully confirm the predictions of our analytical calculations,
namely that the vortex dynamics is the same under both kinds of noise. 
An example of our results is shown in Fig.\ \ref{fig1} for additive 
noise and in Fig.\ \ref{fig2} for multiplicative noise. It is 
already evident from
comparing the dispersion in the $R$ and $\phi$ components that the
behavior of the vortex is very approximately equal in the two problems.

Furthermore, the comparison to the analytical prediction is very good:
We emphasize that there is only one adjustable parameter, 
the vortex diffusion constant, whose value we can only estimate as 
an exact expression for the discrete structure of the vortex core
is not known. The systematic deviation of the 
theory for very long times, $t\gtrsim 2000$, is due to a simplification,
\cite{Till3}
namely keeping $R_0$ constant; in the simulation, $R_0$ slowly 
increases because of the damping. Results for other
low
temperatures compare equally well, the better the lower the temperature;
for higher temperatures, our numerical estimates are less accurate,
although qualitatively the results for both cases remain the same. 
         
\section{Conclusions}

In this work, we have provided analytical and numerical evidence
that additive (Langevin-like) and 
multiplicative noise (coming from fluctuations
in the local field) have the same effects on the dynamics of 
vortices described by the 2D Landau-Lifshitz-Gilbert equation. 
Analytically, the result has been obtained in the framework of a 
collective coordinate approach within a generalized travelling 
wave {\em Ansatz}.\cite{Mertens96} The variances of the effective 
force acting on the vortex were shown to be the same in the additive
and multiplicative cases provided that deformations of the unperturbed
vortex shape can be neglected.
It is important to stress that this 
hypothesis is not an extra assumption but rather it is in fact implicitly 
made when using any collective coordinate approach.\cite{review}
In order to substantiate our analytical results,
we numerically simulated both the additive and the multiplicative
cases for the 2D easy-plane Heisenberg ferromagnet, finding an 
excellent agreement with the prediction of equal behavior under 
both sources of noise, as well as with the analytical expression for 
the variance of the vortex trajectory.

As an immediate consequence of the validity of the above discussed 
prediction, we point out that 
all the results obtained in Ref.\ \onlinecite{Till3} for the additive
(thermal) noise apply to the multiplicative model presented here,
specifically: i) the existence of three
different temperature regimes for the vortex propagation: a low temperature
one, where the vortex motion follows essentially the third order equation
of motion with parameters independent of temperature; a middle temperature
one, at which traces of the oscillations arising from the third order
equation are lost, and a high temperature regime, which is not describable
by a one-vortex approach because too many vortex-antivortex pairs arise
in the system; and ii) the dependence of the effective diffusion 
coefficient for the vortex on temperature. On the other hand, the work
we have reported on here allows to place on firmer grounds that 
those are indeed the features of thermal vortex dynamics: The problem 
with the non-constant spin length in a Langevin approach is now 
solved by the multiplicative approach, in which it is exactly conserved.
Furthermore, the fact that the stationary probability distribution for
the multiplicative case is precisely the Boltzmann factor reinforces 
our conclusion that this is the correct description of thermal effects 
in the framework of models with dynamics given by the 
Landau-Lifshitz-Gilbert equation. The thorough study carried out in 
Ref.\ \onlinecite{Till3} for the Langevin approach is then fully 
confirmed by the present work.

\section*{Acknowledgments} 

We thank Grant Lythe, Yuri Gaididei, 
Francisco Dom\'\i nguez-Adame, Ra\'ul Toral,
and Niels Gr\o nbech-Jensen for discussions. 

Travel 
between Bayreuth and Madrid is supported by ``Acciones
Integradas Hispano-Alemanas'', a joint program of DAAD (Az.\ 314-AI)
and DGES.
Additional funding for
the stay of E.\ M.\ at Bayreuth was provided by 
the Fundaci\'on Universidad Carlos III. 
Travel between Europe and Los Alamos is supported by NATO
grant CRG 971090.
Work at Legan\'es is supported by CICyT (Spain)
grant MAT95-0325 and by DGES (Spain) grant PB96-0119.
Work at Los Alamos is supported by the United States Department of Energy.

\appendix

\section{Noise in the Hamilton equations}

In this section we consider an alternative formulation of the problem of 
thermal fluctuations in the classical
Heisenberg ferromagnet, and show that this 
version exhibits different features for additive
and multiplicative noises, not even being well posed for the latter case. 

By taking into account that the spin length has to be constant, one can 
reformulate the Heisenberg model in terms of the 
fields $\phi=\arctan(S_y/S_x)$ and $\psi=S_z$ for the spin vector: In this
way, after rewriting the Hamiltonian (\ref{eq:Hamiltonian}) in these new
variables, the dynamics of the model is governed by the following 
Hamilton equations: 
\begin{mathletters}
\label{ham}
\begin{eqnarray}
\dot{\phi}&=&\frac{\delta H}{\delta\psi},\\
\dot{\psi}&=&-\frac{\delta H}{\delta\phi}.
\end{eqnarray}
\end{mathletters}
The additive noise version of the problem, i.e., the Langevin equations,
is given by
\begin{mathletters}
\label{hamadd}
\begin{eqnarray}
\dot{\phi}&=&\frac{\delta H}{\delta\psi}+\eta_{\phi}(\vec{r},t),\\
\dot{\psi}&=&-\frac{\delta H}{\delta\phi}+\eta_{\psi}(\vec{r},t),
\end{eqnarray}
\end{mathletters}
whereas the multiplicative version corresponds to 
\begin{mathletters}
\label{hammult}
\begin{eqnarray}
\dot{\phi}&=&\frac{\delta H}{\delta\psi}+\eta_{\phi}(\vec{r},t)\phi,\\
\dot{\psi}&=&-\frac{\delta H}{\delta\phi}+\eta_{\psi}(\vec{r},t)\psi.
\end{eqnarray}
\end{mathletters}
Of course, both models should also be supplemented 
with damping terms in order to fulfill the fluctuation-dissipation 
theorem. For simplicity, we will leave them out in this discussion. 
Their influence, however, is not qualitatively important as for our 
main argument below, as it can be checked that they would only 
contribute factors to the expressions for the force variances. 

Now, from either one of these two equations we can derive an equation 
of motion for the vortex center following basically the same collective 
coordinate approach as above. In this case, one has to multiply the 
equation for $\dot{\psi}$ by $\partial\phi/\partial X_i$ and substract 
from it the equation for $\dot{\phi}$ times $\partial \psi/\partial X_i$. 
In this way an equation completely analogous to the third order 
equation of motion (\ref{eq:3rd-ord_damp}) is obtained, but now we 
find for the forces in the additive case
\begin{equation} \label{eq:FF}
   F^{{\rm add}}_i = \int\!d^2r\, \left[\pdiff{{\cal\phi}}{X_i} \eta_{\psi} -
\pdiff{\psi}{X_i} \eta_{\phi}\right]  ,
\end{equation}
whereas for the multiplicative noise, 
\begin{equation} \label{eq:FFm}
   F^{{\rm mult}}_i = \int\!d^2r\, \left[\pdiff{{\cal\phi}}{X_i}\psi \eta_\psi -
\pdiff{\psi}{X_i} \phi \eta_{\phi}\right]  .
\end{equation}

Let us first discuss the additive case, Eq.\ (\ref{eq:FF}). It is 
not difficult to derive from the above expression the first
moments of the additive force, which turn out to be
$\langle F^{{\rm add}}_i (t)\rangle = 0$ and 
\begin{multline}
\langle F^{{\rm add}}_i (t) F^{{\rm add}}_i (t') \rangle = \\  D \delta(t-t')
\int\!d^2r\,\left[\left(\pdiff{{\cal\phi}}{X_i}\right)^2 +
\left(\pdiff{\psi}{X_i}\right)^2\right]
\label{mas1}
\end{multline}
This expression is actually not very different from that obtained
in Eq.\ (\ref{eq:Fst_var})
for the additive noise version of the Landau-Lifshitz equation:
Using Eq.\ (\ref{eq:S_m_phi}), the integrand of 
Eq.\ (\ref{eq:Fst_var}) reads, for $i=j$,
\begin{equation}
\label{zzz1}
\left(\frac{\partial\vec{S}}{\partial X_i}\right)^2=
(1-\psi^2)\left(\frac{\partial\phi}{\partial X_i}\right)^2+
\frac{1}{1-\psi^2}\left(\frac{\partial\psi}{\partial X_i}\right)^2.
\end{equation}
As the static $S_z$ structure of the vortex (\ref{jolin})
falls off exponentially in 
the outer region $a_c\leq r\leq L$, where $L$ is the system radius,
both integrals (\ref{eq:Fst_var}) and (\ref{mas1}) have the {\em same}
logarithmic
size dependence,
\begin{equation}
\label{zzz2}
\langle F^{{\rm add}}_i (t) F^{{\rm add}}_i (t') \rangle =   D \delta(t-t')
\pi \left(\log\frac{L}{a_c}+C\right).
\end{equation}
The only difference between both formulae consists in the value of the
(small) constant $C$, which stems from the core region $0\leq r \leq a_c$,
where $a_c={\cal O}(r_v)$.

Moving now to the multiplicative case, given by Eq.\ (\ref{eq:FFm}),
it can be easily shown that the situation is completely different. 
The mean and the variance of $F^{{\rm mult}}_i$ can again be easily computed,
yielding $\langle F^{{\rm mult}}_i(t)\rangle=0$ and 
\begin{multline}
\langle F^{{\rm mult}}_i (t) F^{{\rm mult}}_i (t') \rangle = \\  D \delta(t-t')
\int\!d^2r\,\left[\left(\pdiff{{\cal\phi}}{X_i}\right)^2\psi^2 +
\left(\pdiff{\psi}{X_i}\right)^2\phi^2\right],
\label{mas2}
\end{multline}
which, due to the presence of a factor $\psi^2$ in the first term of the 
integrand decaying exponentially
away from the vortex core, is {\em independent} of the system 
size. It is then clear that the variances are very different in both 
cases.
Therefore, a detailed study of the Hamilton equations with
additive noise, Eqs.\ (\ref{hamadd}) would basically lead to the
same conclusions summarized above arising from the study of the 
Landau-Lifshitz-Gilbert problem, whereas the multiplicative version 
would give totally different answers. It has to be concluded then that
the proper way to include multiplicative noise in this problem is 
at the Landau-Lifshitz level and {\em not} in the 
Hamilton equations for $\phi$ and $\psi$. Finally, this result points out very
clearly that the fact that the two approaches, additive and multiplicative,
give the same results in the Landau-Lifshitz equation, is neither 
trivial  nor generally applicable, and should then be regarded as a 
specific and attractive feature of that approach.

\end{multicols}

\begin{references}
\bibitem{nota1} The sign of our
damping term differs from Refs.\ \onlinecite{Thiele73},
\onlinecite{Thiele74}, and \onlinecite{Iida62} because we
work with spins while these authors deal with magnetisations.
\bibitem{nota2} In fact, this result is for the Landau-Lifshitz equation
with Landau damping. However, one can reproduce the calculations in,
e.g., Garanin's paper\cite{garanin} (although they become somewhat more 
involved) to check that the result holds for Gilbert damping as well.
\end{references}
\end{document}